\documentstyle[12pt]{article}
\newcounter{eqn}
\def\lab{\refstepcounter{eqn}\eqno(\arabic{eqn})}
\def\l#1{\lab\label{#1}}
\def\r#1{(\ref{#1})}

\begin{document}

\begin{titlepage}
\begin{flushright}
July 5,1998\\
\end{flushright}

\begin{flushright}
hep-th/9807039\\
\end{flushright}

\begin{centering}
\vfill
{\bf RENORMALIZATION  OF  SPATIALLY INHOMOGENEOUS NONEQUILIBRIUM FIELD
DYNAMICS}

\vspace{1cm}
O. Yu. Shvedov \footnote{olshv@ms2.inr.ac.ru, shvedov@qs.phys.msu.su} \\
\vspace{0.3cm}
{\small {\em Chair of Quantum Statistics and Field Theory,}}\\
{\small{\em Department of Physics, Moscow State University }}\\
{\small{\em Vorobievy gory, Moscow 119899, Russia}}

\vspace{0.7cm}

{\bf Abstract}

\end{centering}
\vspace{0.7cm}

The problem of renormalization of the semiclassical one-loop equations
used in the non-equilibrium field theory is considered.  Recently, the
renormalizability of  such  equations  has  been  justified  for  some
partial cases of classical field configurations.  In  this  paper  the
general case  of arbitrary spatially inhomogeneous field configuration
is investigated.  It is shown that  for  certain  quantum  states  the
divergences arising  in  one-loop equations can be eliminated by usual
perturbation-theory counterterms.

\vfill \vfill
\noindent

\end{titlepage}
\newpage

There are different approaches to the nonequilibrium field theory. One
can  consider  various types of the semiclassical technique \cite{SC},
the   one-loop   approximation   \cite{B,B1},    the    time-dependent
Hartree-Fock  method  \cite{B,HF}  based  on  the variational Gaussian
approach \cite{G},  the large-$N$ nonequilibrium  equations  \cite{1N}
based  on  the $1/N$-expansion technique \cite{1N1}.  These approaches
play an important role in quantum cosmology,  as well as  in  particle
physics.

Usually, only homogeneous mean field  configurations  are  taken  into
account. For   this  case,  the  problem  of  renormalization  of  the
nonequilibrium dynamics was considered in details \cite{HF,B,B1}.

However, the  processes  with  inhomogeneous condensates are also very
important \cite{B98},  so that it  is  necessary  to  investigate  the
problem  of  renormalization of the inhomogeneous nonequilibrium field
theory.  Some examples of such a renormalization are  investigated  in
ref.\cite{B98}.

In this  paper  the  renormalization  procedure  is considered for the
general inhomogeneous field configurations.
It will be shown that it is sufficient to
use  the ususal perturbation-theory terms only in order to renormalize
the one-loop equations.

Let us  consider  the $\Phi^4$-theory,  the simplest (3+1)-dimensional
renormalizable model with the Lagrangian:
$$
{\cal L} = \frac{1+\delta Z}{2} (\partial_{\mu}\varphi)^2-
\frac{m^2+\delta m^2}{2} \varphi^2
-\frac{\lambda+\delta\lambda}{24} \varphi^4
\l{e1}
$$
depending on  the  small  parameter  $\lambda$  being   the   coupling
constant. Expression  \r{e1} contains the counerterms being calculable
with the help of the perturbation theory,
$$
\delta Z \sim \lambda^2,  \delta m^2 \sim \lambda, \delta \lambda \sim
\lambda^2.
\l{e1a}
$$
A usual  way  to  derive  the one-loop nonequilibrium equations
\cite{B} is the
following. One starts from the Heisenberg equations of motion
$$
(1+\delta Z) \partial_{\mu} \partial_{\mu} \hat{\varphi} +
(m^2+\delta m^2) \hat{\varphi} + \frac{\lambda+\delta\lambda}{6}
\hat{\varphi}^3 =0
\l{e2}
$$
on the Heisenberg field operator $\hat{\varphi}(x)$.  The next step is
to consider  the  special  (``semiclassical'')  quantum  states.   One
supposes that  the average values of the quantum field $\hat{\varphi}$
is of order $O(1/\sqrt{\lambda})$,
$
<\hat{\varphi}> = \frac{\Phi}{\sqrt{\lambda}},
%\l{e2*}
$
while the  average values of all regular in $\lambda$ functions of the
difference $\hat{\varphi} - <\hat{\varphi}>$ are of order  $O(1)$.  This
choice of  the  quantum  sattes  corresponds to the Maslov complex-WKB
theory \cite{M}.
Making use of this conjecture,  one can decompose  the  quantum  field
$\hat{\varphi}$ into classical and quantum terms,
$$
\hat{\varphi} = \frac{\Phi}{\sqrt{\lambda}} + \hat{\psi},\qquad
<\hat{\psi}>=0
$$
Substituting this relation to  eq.\r{e2},  one
obtains
$$
(1+\delta Z)\partial_{\mu}\partial_{\mu}
(\Phi+\sqrt{\lambda}\hat{\psi}) + (m^2+\delta m^2)
(\Phi+\sqrt{\lambda}\hat{\psi}) +
\frac{1+\delta\lambda/\lambda}{6}
(\Phi+\sqrt{\lambda}\hat{\psi})^3 =0
\l{e4}
$$
To find the leading order of the mean field $\Phi$, one should perform
a limit $\lambda\to 0$ in eq.\r{e4}. Using estimations \r{e1a}, one
obtains the classical equations of motion,
$
\partial_{\mu}\partial_{\mu} \Phi + m^2\Phi + \frac{\Phi^3}{6} =0.
%\l{e4*}
$
To study the next order of the mean field, $O(\lambda)$, one can
average eq.\r{e4}, make use of eqs.\r{e1a}, identity $<\hat{\psi}>=0$
and take into account only the terms of orders $O(1)$ and $O(\lambda)$,
$$
\partial_{\mu}\partial_{\mu}\Phi +
(m^2+\delta m^2) \Phi + \frac{1+\delta \lambda/\lambda}{6}\Phi^3+
\frac{\lambda}{2}\Phi <\hat{\psi}^2>=0.
\l{e5}
$$
It is also necessary to find a leading order approximation for
$\hat{\psi}$. Considering terms of order $O(\sqrt{\lambda})$ in
eq.\r{e4} and taking into account the classical equation, one
finds the following equation of motion
$$
\partial_{\mu} \partial_{\mu} \hat{\psi}
+ (m^2+\frac{1}{2}\Phi^2) \hat{\psi} =0.
\l{e6}
$$
The set \r{e5}, \r{e6} called as one-loop equations is to be
investigated.

For the case of spatially homogeneous mean field configurations, it is
very iuseful to present the field operator $\hat{\psi}$ as a linear
superposition of creation and annihilation operators,
$
\hat{\psi}(t,{\bf x}) = \int \frac{d{\bf
k}}{(2\pi)^{3/2}\sqrt{2\omega_{\bf k}} }
[a^-({\bf k})U_{\bf k}(t)e^{i{\bf k}{\bf x}}
+
a^+({\bf k})U^*_{\bf k}(t)e^{-i{\bf k}{\bf x}}],
%\l{e7}
$
$\omega_{\bf k}=\sqrt{{\bf k}^2+m^2}$, since the equations on $U_{\bf
k}(t)$ are independent for different values of the momentum $\bf k$.
For spatially inhomogeneous configurations, this ansatz  does not
obey eq.\r{e6}, so that it is necessary to consider the functions
$U_{\bf k}(t)$ to be $\bf x$-dependent,
$$
\hat{\psi}(t,{\bf x}) = \int \frac{d{\bf
k}}{(2\pi)^{3/2}\sqrt{2\omega_{\bf k}}}
[a^-({\bf k})U_{\bf k}({\bf x},t)e^{i{\bf k}{\bf x}}
+
a^+({\bf k})U^*_{\bf k}({\bf x},t)e^{-i{\bf k}{\bf x}}].
\l{e8}
$$
The quantum-field equation \r{e6} implies the following relation on
the coefficient functions $U_{\bf k}({\bf x},t)$:
$$
\frac{\partial^2}{\partial t^2}U_{\bf k}({\bf x},t) +
\left[
\left(
{\bf k}- i \frac{\partial}{\partial {\bf x}}
\right)^2
+m^2+\frac{1}{2}f({\bf x},t)
\right]
U_{\bf k}({\bf x},t)=0.
\l{e9}
$$
where
$$
f({\bf x},t)=\Phi^2({\bf x},t).
%\l{e9*}
$$
Since the transformation \r{e8} should lead to usual canonical
commutation relations,
$$
[a^{\pm}({\bf k}), a^{\pm}({\bf p})]=0,
[a^-({\bf k}),a^+({\bf p})]=\delta({\bf k}-{\bf p}),
\l{e10}
$$
it is necessary to impose the following constraints on $U_{\bf k}({\bf
x},t)$:
$$
\begin{array}{c}
\int \frac{d{\bf k}}{(2\pi)^32\omega_{\bf k}}
[ U_{\bf k}({\bf x},t) U_{\bf k}^*({\bf y},t)
- U_{-{\bf k}}^*({\bf x},t) U_{-{\bf k}}({\bf y},t) ]
e^{i{\bf k}({\bf x}-{\bf y})} = 0.\\
\int \frac{d{\bf k}}{(2\pi)^32\omega_{\bf k}}
[ \dot{U}_{\bf k}({\bf x},t) \dot{U}_{\bf k}^*({\bf y},t)
- \dot{U}_{-{\bf k}}^*({\bf x},t) \dot{U}_{-{\bf k}}({\bf y},t) ]
e^{i{\bf k}({\bf x}-{\bf y})} = 0.\\
\int \frac{d{\bf k}}{(2\pi)^32\omega_{\bf k}}
[ U_{\bf k}({\bf x},t) \dot{U}_{\bf k}^*({\bf y},t)
- U_{-{\bf k}}^*({\bf x},t) \dot{U}_{-{\bf k}}({\bf y},t) ]
e^{i{\bf k}({\bf x}-{\bf y})} = 0.
\end{array}
\l{e11}
$$
being corollaries of the canonical commutation relations on the field
$\hat{\psi}({\bf x},t)$ and eqs.\r{e10}. The conditions \r{e11} are
invariant under time evolution, so that it is sufficient to impose
them at the initial time moment only.
The average value $<\hat{\psi}^2>$ can be presented
as a sum of  vacuum and regular parts:
$
<\hat{\psi}^2> = \int \frac{d{\bf k}}{(2\pi)^3 2\omega_{\bf k}}
|U_{\bf k}({\bf x},t)|^2 + <:\hat{\psi}^2:>.
$
where $<:\hat{\psi}^2:>$ is finite. This implies that eq.\r{e5}
takes the form
$$
\partial_{\mu}\partial_{\mu}\Phi +
(m^2+\delta m^2) \Phi + \frac{1+\delta \lambda/\lambda}{6}\Phi^3
+\frac{\lambda}{2}\Phi
\int \frac{d{\bf k}}{(2\pi)^3 2\omega_{\bf k}}
|U_{\bf k}({\bf x},t)|^2
+ \frac{\lambda}{2}\Phi <:\hat{\psi}^2:>=0.
\l{e13}
$$
To investigate the divergences of \r{e13}, it is necessary to
investigate the asymptotic behaviour of $U_{\bf k}({\bf x},t)$ at
large values of $|{\bf k}|$ up to $O(\omega_{\bf k}^{-2})$.

There are several ways to construct the approximate solution to
eq.\r{e9} at large values of $|{\bf k}|$.
For example, due to the WKB-approach
\cite{M3}, one can consider the substitution
$
U_{\bf k}=
e^{i\omega_{\bf k}t}V_{\bf k}^+
+
e^{-i\omega_{\bf k}t}V_{\bf k}^-
$
and consider the perturbation theory for  $V_{\bf  k}^{\pm}$.  Another
approach  is  the following.  One can apply the perturbation theory in
$\Phi^2$ \cite{B1}.  Expanding the function  $U_{\bf  k}$,
$$
U_{\bf
k}=U_{\bf   k}^0   +   U_{\bf   k}^1   +   U_{\bf  k}^2  +...,
\qquad
U_{\bf k}^0=e^{-i\omega_{\bf k}t},
$$
one finds the  following  set  of  the
recursive   relations,
$$   \frac{\partial^2}{\partial  t^2}U^m_{\bf
k}({\bf x},t) + \left[ \left( {\bf k}- i \frac{\partial}{\partial {\bf
x}}  \right)^2  +m^2  \right]  U_{\bf k}^m({\bf x},t)=j_{\bf k}^m({\bf
x},t).
\l{e14}
$$
where
$$      j_{\bf      k}^m({\bf
x},t)=-\frac{1}{2}f({\bf    x},t)U^{m-1}_{\bf    k}({\bf   x},t).
%\l{e14*}
$$
There is a problem of imposing initial  conditions  \cite{B2}  on  the
functions $U_{\bf  k}$  obeying eqs.\r{e9}.  The simplest choice is to
impose a free theory initial conditions:
$$
U^m_{\bf k}({\bf
x},0)=0,  \dot{U}^m_{\bf  k}({\bf  x},0)=0,  m\ge  1
$$
The functions $U_{\bf k}^m$  can  be
expressed then  via the right-hand side of eq.\r{e14}:
$$
\begin{array}{c}
U^m_{\bf k}({\bf
x},t)= U^{m+}_{\bf k}({\bf x},t)+ U^{m-}_{\bf k}({\bf x},t),\\
U^{m\pm}_{\bf k}({\bf x},t) = \int_{t_0}^t d\tau \int \frac{d{\bf
p}d{\bf  y}}{(2\pi)^3}  e^{i{\bf  p}({\bf  x}-{\bf  y})}  \frac{e^{\pm
i\omega_{{\bf  k}+{\bf  p}}(t-\tau)}} {\pm 2i\omega_{{\bf k}+{\bf p}}}
j^m_{\bf k}({\bf y},\tau)
\end{array}
\l{e15}
$$
for
$
t_0=0.
$
The integral over $\bf  p$  and
$\bf y$ can be taken by the stationary phase technique
(see, for example, \cite{M3}).
The stationary  points  of  the  phase  ${\bf  p}({\bf x}-{\bf y}) \pm
\omega_{{\bf k}+{\bf p}}(t-\tau)$ are
$$
{\bf p}=0,  {\bf y}_{\pm} = {\bf x} \pm \frac{\bf  k}{\omega_{\bf  k}}
(t-\tau),
\l{e15*}
$$
so that   one   should   perform   a   substitution   ${\bf    p}={\bf
s}\sqrt{\omega_{\bf k}}$,    ${\bf    y}={\bf    y}_{\pm}    +    {\bf
z}/\sqrt{\omega_{\bf k}}$ and expand the
integrand in eq.\r{e15} into a series in $\omega_{\bf k}^{-1/2}$.

However, to make the technical calculations easier, one can perform
the following operational trick
(cf.\cite{M3}) being equivalent to the direct
stationary phase calculations. One can represent eq.\r{e15} as follows,
$$
U^{m\pm}_{\bf k}({\bf x},t) =
\int_{t_0}^t d\tau \frac{e^{\pm i\omega_{\bf k}(t-\tau)}}
{\pm 2i\omega_{\bf k}}
\int \frac{d{\bf p}d{\bf y}}{(2\pi)^3}
\xi_{{\bf k},t-\tau}({\bf p})
j^m_{\bf k}(
{\bf x} \pm \frac{\bf k}{\omega_{\bf k}}(t-\tau)
+i\frac{\partial}{\partial {\bf p}},\tau)
e^{i{\bf p}({\bf x}-{\bf y}\pm \frac{\bf k}{\omega_{\bf k}}(t-\tau))},
\l{e16}
$$
where
$$
\xi_{{\bf k},t-\tau}({\bf p})=
\frac{\omega_{\bf k}}{\omega_{{\bf k}+{\bf p}}}
e^{\pm i (\omega_{{\bf k}+{\bf p}}-\omega_{\bf k}
- \frac{{\bf k}{\bf p}}{\omega_{\bf k}}) (t-\tau)}
$$
since the function
$
e^{i{\bf p}({\bf x}-{\bf y}\pm \frac{\bf k}{\omega_{\bf k}}(t-\tau))}
$
is an eigenfunction of the operator
$-i\frac{\partial}{\partial {\bf p}}$ which corresponds to the
eqigenvalue
${\bf x}-{\bf y}\pm \frac{\bf k}{\omega_{\bf k}}(t-\tau))$. One should
then integrate eq.\r{e16} over ${\bf p}$ by parts,
$$
U^{m\pm}_{\bf k}({\bf x},t) =
\int_{t_0}^t d\tau \frac{e^{\pm i\omega_{\bf k}(t-\tau)}}
{\pm 2i\omega_{\bf k}}
\int \frac{d{\bf p}d{\bf y}}{(2\pi)^3}
e^{i{\bf p}({\bf x}-{\bf y}\pm \frac{\bf k}{\omega_{\bf k}}(t-\tau))}
j^m_{\bf k}(
{\bf x} \pm \frac{\bf k}{\omega_{\bf k}}(t-\tau)
-i\frac{\partial}{\partial {\bf p}},\tau)
\xi_{{\bf k},t-\tau}({\bf p}),
$$
where the function $\xi$
can be expanded in $1/\omega_{\bf k}$:
$$
\xi_{\bf k}({\bf p},t-\tau)=
1
- \frac{{\bf p}{\bf k}}{\omega^2_{\bf k}}
\pm
\frac{i(t-\tau)}{2\omega_{\bf k}}
\left[{\bf p}^2 - \frac{({\bf p}{\bf k})^2}{\omega_{\bf k}^2}\right]
+ O(\omega_{\bf k}^{-2}).
\l{e16a}
$$
Since the integration over ${\bf y}$ gives us a $\delta$-function
$\delta({\bf p})$, one finds the following expression:
$$
U^{m\pm}({\bf x},t) =
\int_{t_0}^t d\tau \frac{e^{\pm i\omega_{\bf k}(t-\tau)}}
{\pm 2i\omega_{\bf k}}
j^m_{\bf k}\left(
{\bf x}\pm \frac{\bf k}{\omega_{\bf k}}(t-\tau)
-i\frac{\partial}{\partial {\bf p}},\tau\right)
\xi_{{\bf k},t-\tau}({\bf p})|_{{\bf p}=0}.
%\l{e17}
$$
Expanding the operator $j^m_{\bf k}$ into a series in
$-i\frac{\partial}{\partial {\bf p}}$,  making use of eq.\r{e16a}, one
finds that
$$
\begin{array}{c}
U^{m\pm}({\bf x},t) =
\int_{t_0}^t d\tau \frac{e^{\pm i\omega_{\bf k}(t-\tau)}}
{\pm 2i\omega_{\bf k}}
\left[
j^m_{\bf k}({\bf y}_{\pm},\tau)
\right.
\\
\left.
+
\left(i \frac{\bf  k}{\omega_{\bf  k}^2}\frac{\partial}{\partial  {\bf
y}_{\pm}}
\mp
\frac{i(t-\tau)}{2\omega_{\bf k}}
\left(
\delta_{jn} - \frac{k_jk_n}{\omega_{\bf k}^2}
\right)
\frac{\partial^2}{\partial y_{\pm}^j \partial y_{\pm}^n}
\right)
j^m_{\bf k}({\bf y}_{\pm},\tau)
+O(\omega_{\bf k}^{-2}) \right],
\end{array}
\l{e18}
$$
where we sum over repeated indeices, while ${\bf y}_{\pm}$ has the form
\r{e15*}.

Eq.\r{e18} allows us to find the functions $U_{\bf k}({\bf  x},t)$  at
larges $|{\bf  k}|$.  It  follows  from  the  first-order perturbation
theory that the function $U^{1-}_{\bf k}({\bf x},t)$ can be expanded as
$$
U^{1-}_{\bf k}({\bf x},t)= e^{-i\omega_{\bf k}t}
\left\{
\int_{t_0}^t d\tau \frac{1}{4i\omega_{\bf k}}
f({\bf x}-\frac{{\bf k}}{\omega_{\bf k}}(t-\tau),\tau)-
\omega_{\bf k}\frac{\partial {\bf g}_{\bf k}}{\partial {\bf k}}
+O(\omega_{\bf k}^{-3})
\right\},
$$
where
$$
{\bf g}_{\bf k}({\bf x},t) = \int_{t_0}^t d\tau \frac{1}{8\omega_{\bf k}^2}
\frac{\partial}{\partial {\bf y}_-} f(
{\bf x}-\frac{{\bf k}}{\omega_{\bf k}}(t-\tau),\tau).
$$
The function $U_{\bf k}^{1+}({\bf x},t)$  presented  as  the  integral
over $t$ of the rapidly oscillating function
$$
U^{1+}_{{\bf k}}({\bf x},t) =
 e^{i\omega_{\bf k}t}
\int_{t_0}^t d\tau
e^{-2i\omega_{\bf k}\tau}
\frac{1}{4i\omega_{\bf k}}
\left[f({\bf x}+\frac{{\bf k}}{\omega_{\bf k}}(t-\tau),\tau)
+O(\omega_{\bf k}^{-1})\right]
$$
is of  order  $O(\omega_{\bf  k}^{-2})$  because  it  is  necessary to
intagrate the expression by parts,
$$
U^{1+}_{\bf k}({\bf x},t) =
e^{i\omega_{\bf k}t}
\left[
-\frac{1}{8\omega_{\bf k}^2}
f({\bf x}+\frac{{\bf k}}{\omega_{\bf k}}(t-\tau),\tau)
e^{-2i\omega_{\bf k}\tau}|^t_{t_0}
+O(\omega_{\bf k}^{-3})
\right].
$$
The leading  order  for  $U^2_{\bf  k}({\bf  x},t)$  is  saturated  by
$U^{2-}_{\bf k}$, $U_{\bf k}^2=U_{\bf k}^{2-}+O(\omega_{\bf k}^{-3})$,
so that
$$
U_{\bf k}^2({\bf x},T) =-\frac{e^{-i\omega_{\bf k}T}}{16\omega_{\bf k}^2}
\int_{t_0}^T dt
f({\bf x}-\frac{\bf k}{\omega_{\bf k}}(T-t),t)
\int_{t_0}^t d\tau
\Phi^2({\bf x}-\frac{\bf k}{\omega_{\bf k}}(T-\tau),\tau)
+O(\omega_{\bf k}^{-3}).
$$
Combining all the terms,  one finds the following expression  for  the
solution to eq.\r{e9} at larges $\omega_{\bf k}$:
$$
\begin{array}{c}
U_{\bf k}({\bf x},t)= \exp[-i\omega_{\bf k}t +
\int_{t_0}^t d\tau  \frac{1}{4i\omega_{\bf   k}}f({\bf   x}-\frac{\bf
k}{\omega_{\bf k}}(t-\tau),\tau)]
[1-\omega_{\bf k}\frac{\partial {\bf g}_{\bf  k}({\bf  x},t)}{\partial
{\bf k}} - \frac{1}{8\omega_{\bf k}^2}f({\bf x},t)]\\
+\frac{1}{8\omega_{\bf k}^2}f({\bf  x}+\frac{\bf   k}{\omega_{\bf
k}}t,t_0) e^{i\omega_{\bf k}(t-2t_0)} + O(\omega_{\bf k}^{-3}),
\end{array}
\l{e19}
$$
where $f=\Phi^2$.

Since the integral of the full derivative vanishes, the divergent part
of the fluctuation integral
$\int \frac{d{\bf k}}{2\omega_{\bf k}} |U_{\bf k}({\bf x},t)|^2$
can be written as
$$
\int \frac{d{\bf k}}{2\omega_{\bf k}}
-\int \frac{d{\bf k}}{8\omega_{\bf k}^3}\Phi^2({\bf x},t)+
\int \frac{d{\bf k}}{8\omega_{\bf k}^3}
\Phi^2({\bf x}+\frac{{\bf  k}}{\omega_{\bf  k}}t,0)  \cos(2\omega_{\bf
k}t)
\l{e19*}
$$
One can  notice  that the divergent parts of the counterterms entering
to eq.\r{e13} should be the following:
$$
\delta m^2 = -\frac{\lambda}{2 (2\pi)^3} 
\int \frac{d{\bf k}}{2\omega_{\bf k}},
\qquad \delta \lambda = \frac{3\lambda^2}{(2\pi)^3}
\int \frac{d{\bf k}}{8\omega_{\bf k}^3},\qquad t\ne 0,
\l{e20}
$$
$$
\delta m^2 = -\frac{\lambda}{2(2\pi)^3}  
\int \frac{d{\bf k}}{2\omega_{\bf k}},
\qquad
\delta \lambda = 0, \qquad t= 0.
\l{e21}
$$
However, the counterterms should be  defined  uniquely,  so  that  one
finds that  there  is  a  singularity  at  the  initial moment of time
\cite{B2}.

This means that another choice of initial conditions for the functions
$U_{\bf k}$ obeying relations \r{e11} and eq.\r{e9} is  necessary.  To
construct such  functions  $U_{\bf  k}$,  one  can  use the Bogoliubov
procedure of switching on the interaction  \cite{BS}.  Choose  $f$  in
eq.\r{e9} to be a smooth function obeying the following properties:
$$
f({\bf x},t) = \left\{
\begin{array}{c}
\Phi^2({\bf x},t), \qquad t\ge t_1  \qquad (t_1<0)
\\
0, \qquad t\ge t_2  \qquad (t_2<t_1)
\end{array}
\right.
$$
instead of $\Phi^2$.
Consider the solution to eq.\r{e9} satisfying the following conditions:
$$
U_{\bf k}({\bf x},t_2) = U^0_{\bf k}({\bf x},t_2),
\quad
\dot{U}_{\bf k}({\bf x},t_2) = \dot{U}^0_{\bf k}({\bf x},t_2),
$$
The canonical relations \r{e11} are satisfied at
$t_0=t_2$ and at arbitrary moment of time as well.  Original  equation
with $f=\Phi^2$  is also satisfied at $t>0$.  The asymptotic expansion
for the solution $U_{\bf k}$ is given  by  eq.\r{e19}  for  $t_0=t_2$.
Since $f({\bf x},t_0)=0$,  while $f=\Phi^2$ at $t>0$, the latter terms
in eq.\r{e19} vanishes, so that the fluctuation integral has the form
$$
\int \frac{d{\bf k}}{2\omega_{\bf k}} |U_{\bf k}({\bf x},t)|^2
= \int \frac{d{\bf k}}{2\omega_{\bf k}}
-\int \frac{d{\bf k}}{8\omega_{\bf k}^3}\Phi^2({\bf x},t)+
\mbox{finite terms}.
$$
The counterterms  \r{e20} calculable in the dimensional regularization
\cite{B2} are sufficient then for eliminating all the divergences  for
this choice of the solution to eq.\r{e9}.

We see  that  there  exists  such a decomposition of the quantum field
$\hat{\psi}$ into creation and annihilation parts  that  provides  the
renormalizability of  the  one-loop  equation \r{e5}.  The form of the
classical field $\Phi$ is not  important  for  performed  derivations,
since it  was not necessary to find the eigenfunctions of eq.\r{e9} as
in \cite{B98}.

The author is indebted to Ju.Baacke,  F.Cooper,  R.Jackiw,  V.P.Maslov
and \\
D.V.Shirkov  for  helpful discussions.  This work was supported by
the Russian Foundation for Basic Research, project 96-01-01544.

\end{document}